\documentclass[conference]{IEEEtran}
\IEEEoverridecommandlockouts
\usepackage{hyperref}
\usepackage{cite}
\usepackage{makecell}
\usepackage{amsmath,amssymb,amsfonts}
\usepackage{algorithmic}
\usepackage{graphicx}
\usepackage{textcomp}
\usepackage{xcolor}
\usepackage{url}
\usepackage{multirow}
\usepackage{booktabs}
\usepackage{booktabs}
\usepackage{graphicx}
\usepackage{dblfloatfix}
\usepackage[normalem]{ulem}
\usepackage{tikz}
\def\BibTeX{{\rm B\kern-.05em{\sc i\kern-.025em b}\kern-.08em
    T\kern-.1667em\lower.7ex\hbox{E}\kern-.125emX}}

\def\sectionautorefname~#1\null{\S#1\null}

\IEEEpubid{%
\begin{minipage}{\textwidth}
    \ \\[50pt]
    \footnotesize
    This work has been submitted to the IEEE for possible publication.
    Copyright may be transferred without notice, after which this version
    may no longer be accessible.
\end{minipage}
}
\begin{document}

\title{Multimodal Speaker Verification as a Threat to Speaker Anonymization
}

\author{
  \IEEEauthorblockN{Ashi Garg, Cristina Aggazzotti, Leibny Paola García-Perera, Nicholas Andrews}
  \IEEEauthorblockA{
    \textit{Johns Hopkins University, Human Language Technology Center of Excellence} \\
     Baltimore, MD, USA \
  }
}
\maketitle
\begin{abstract}
Most automatic speaker verification (ASV) systems operate on individual utterances, despite real-world interactions typically consisting of multiple utterances. As speech accumulates, increasingly rich speaker information becomes available through acoustic, prosodic, and linguistic cues, potentially challenging speaker anonymization methods that primarily target vocal characteristics. We investigate ASV in a multi-utterance, multimodal setting and examine whether aggregating information across anonymized speech impacts privacy. We first study audio-only aggregation across multiple anonymized utterances and observe consistent performance improvements as more speech becomes available. We then incorporate prosodic and linguistic information, showing that multimodal systems outperform unimodal approaches. Finally, we compare aggregation strategies and find that frame-level aggregation yields the lowest EERs. Even with only five anonymized utterances, combining audio and text reduces EER by over 15\% relative to audio-only aggregation, demonstrating that substantial speaker-discriminative information remains accessible despite anonymization.

\end{abstract}


\section{Introduction}

Speech recordings convey far more than their spoken message: they encode a speaker's vocal characteristics, prosodic habits, and linguistic choices, all of which can be exploited to infer the speaker's identity. Speaker anonymization seeks to mitigate this risk by transforming speech so that the speaker can no longer be recognized while the linguistic content remains intact, and standardized evaluation of such systems has been established through the VoicePrivacy Challenge~\cite{tomashenko2024voiceprivacy}. However, existing anonymization methods and their evaluation protocols operate predominantly on individual, isolated utterances and focus on suppressing a speaker's \emph{acoustic} identity.

In contrast, much of human spoken communication occurs across multiple utterances, such as conversations and interviews. As speech accumulates, increasingly rich information about a speaker becomes available, including vocal traits, prosodic patterns such as pitch and speaking rate, and linguistic content reflecting an individual's vocabulary and syntax. An attacker with access to several utterances from the same speaker may therefore aggregate speaker-identifying information that no single utterance reveals, challenging anonymization methods designed around per-utterance acoustic transformation. Nevertheless, automatic speaker verification (ASV) systems, including those used as attack models in privacy evaluations, generally process isolated utterances, where speaker representations are extracted from individual, short utterances and compared to obtain a verification score~\cite{tomashenko2024voiceprivacy}.

In this work, we investigate multi-utterance speaker verification, where information from several utterances is aggregated to construct more robust speaker representations\footnote{Code and pretrained models are available at \url{https://github.com/Ashigarg123/multimodal-speaker-verification}.}. Specifically, we explore various aggregation strategies at both the utterance level and fine-grained frame level, allowing the system to capture speaker characteristics at different temporal resolutions.

Besides acoustic information, speaker properties can also be revealed from additional speaker-discriminative cues in the speech signal~\cite{aggazzottigarg2026}. These include linguistic content derived from automatic speech recognition (ASR) transcripts~\cite{aggazzotti2024, aggazzotti2025asr} and prosodic features. We thus explore multimodal speaker representations that combine acoustic, linguistic, and prosodic information and investigate how these cues can be effectively aggregated across multiple utterances. Finally, we assess the robustness of the proposed aggregation approaches in the challenging setting of anonymized speech generated through voice conversion~\cite{kuzmin2026}, where the speaker's acoustic identity is partially suppressed. Adopting the attacker's perspective of the VoicePrivacy framework, we consider both a lazy-informed attacker, trained only on original speech, and a stronger semi-informed attacker with access to anonymized training data.

We focus on three main research questions:
\begin{itemize}
    \item \textbf{RQ1}: Does aggregating audio over multiple utterances improve ASV performance on \emph{anonymized speech}?
    \item \textbf{RQ2}: Does incorporating complementary modalities, such as linguistic content and prosodic signals, into multimodal systems outperform unimodal systems?
    \item \textbf{RQ3}: Is utterance-level or frame-level aggregation more effective for multimodal speaker verification?
\end{itemize}

Our experiments on conversational telephone speech yield three main findings. First, aggregating audio information across multiple anonymized utterances consistently improves ASV performance, with gains that grow as more utterances become available. Second, multimodal systems that combine acoustic embeddings with prosodic features and linguistic content extracted from ASR transcripts outperform their unimodal counterparts, indicating that linguistic and prosodic cues retain substantial speaker-identifying information after voice anonymization. Third, comparing aggregation strategies reveals that frame-level aggregation is consistently the most effective, suggesting that speaker-discriminative information preserved after anonymization is distributed across temporal regions and is best exploited before utterance-level pooling. Taken together, these results demonstrate that combining multiple utterances with complementary modalities reveals residual speaker-identifying information, reducing speaker privacy even under challenging anonymization conditions and motivating anonymization methods and evaluation protocols that account for multi-utterance, multimodal attackers.

\section{Related Work}
The VoicePrivacy Challenge aims to obscure a speaker's acoustic identity by anonymizing their voice in individual, isolated utterances~\cite{tomashenko2024voiceprivacy}. However, when multiple utterances from the same speaker are available, speaker identity may still be inferred from information accumulated across utterances, including linguistic patterns, such as an individual's vocabulary and syntax~\cite{aggazzottigarg2026}. Thus, voice anonymization alone may not be sufficient to fully suppress speaker-identifying information. 

Previous work has explored multi-utterance aggregation strategies, particularly to combine information across multiple enrollment utterances \cite{zeng2022attention, krzywdziak2025merge}. While attention-based and statistical aggregation strategies have been proposed, their effectiveness has primarily been evaluated on original speech. Their applicability to anonymized speech and the resulting impact on speaker privacy remain largely unexplored. Furthermore, it remains unclear which aggregation strategies are most effective for accumulating speaker-discriminative information across multiple utterances.

Speaker identity may be encoded across multiple sources of information beyond conventional acoustic embeddings. 
In addition to linguistic cues, other speech attributes, such as prosody~\cite{mary2006prosodic,shriberg2005modeling,soleymani2018prosodic}, fundamental frequency~\cite{vandommelen1990}, %speaking rate~\cite{}, 
and rhythm~\cite{mehlman2025rhythm}, have been shown to encode speaker-discriminative information. Other work has explored combining facial and speech information for audio-visual person verification \cite{praveen2024audio,chen20253d}, demonstrating the benefits of leveraging complementary information sources. 

In summary, these findings suggest that speaker identity may be encoded across multiple sources of information, including acoustic, linguistic, prosodic, and other complementary cues. However, existing work has primarily focused on speaker recognition and verification in original, unanonymized speech. While voice anonymization does not perfectly disentangle speaker identity from other speech attributes~\cite{shamsabadi2022differentially}, it remains unclear which of these cues continue to encode speaker-identifying information after anonymization and to what extent they can be exploited for speaker recognition.

\section{Proposed Method}
In this section, we present the audio-only aggregation methods, followed by the proposed multimodal aggregation. We investigate aggregation at two levels. First, we study \emph{utterance-level aggregation}, where each utterance is independently encoded and aggregation is performed over utterance embeddings. Second, we explore \emph{frame-level aggregation}, where frame representations are combined before speaker pooling to capture fine-grained temporal information. For both settings, we examine audio-only as well as multimodal aggregation strategies incorporating linguistic and prosodic information.

\subsection{Utterance-level Aggregation}

Aggregation methods operating on utterance-level embeddings are extracted independently from each utterance.

\subsubsection{Audio Query Attention}\label{query_attention}
Utterances from the same speaker may contain varying amounts of speaker-discriminative information due to differences in phonetic content, speaking style, and recording conditions. We therefore employ a learnable query attention mechanism to adaptively weight utterances during multi-utterance aggregation.

Let $\mathbf{X}
=
[\mathbf{x}_1,\mathbf{x}_2,\ldots,\mathbf{x}_N]
\in \mathbb{R}^{N\times D}$ denote the set of utterance embeddings associated with a speaker, where $N$ is the number of utterances, $\mathbf{x}_i \in \mathbb{R}^{D}$, and $D$ is the embedding dimension. A learnable query vector $\mathbf{q}\in\mathbb{R}^{D}$ is used to attend over the utterance embeddings through multi-head attention
\[
Q=\frac{\mathbf{q}}{\tau},
\qquad
K=V=\mathbf{X},
% \]
% \[
\qquad 
\mathbf{z}
=
\mathrm{MHA}(Q,K,V),
\] 
where $\tau$ is a temperature parameter and $\mathbf{z}\in\mathbb{R}^{D}$ denotes the aggregated speaker embedding.

\subsubsection{Utterance-level Audio-Text Fusion}
\label{multimodal_audio_text}
While speaker anonymization primarily targets acoustic characteristics, linguistic content may retain speaker-discriminative information. We therefore fuse audio and text embeddings at the utterance level to capture complementary speaker information across modalities.

For each utterance, we extract an audio embedding using WavLM-ECAPA-TDNN \cite{desplanques2020ecapa,chen2022wavlm} and a text embedding using the authorship representation LUAR \cite{uar-emnlp2021}. These models are described in \autoref{sec:baseline}. The two embeddings are projected into a shared 256-dimensional space and layer normalized. The projected embeddings are concatenated and subsequently mapped to a 192-dimensional multimodal speaker representation through a linear projection layer.

\subsubsection{Utterance-level Audio-Prosody Fusion}
\label{multimodal_prosody}

Prosodic characteristics, such as pitch and speaking rate, provide complementary 
speaker information beyond conventional speaker embeddings. We therefore 
augment the acoustic representation with three utterance-level prosodic 
features: mean fundamental frequency (\(F0_{\text{mean}}\)), voiced ratio 
(\(r_{\text{voiced}}\)), and speaking rate (\(r_{\text{spkrate}}\)).

\paragraph{Prosodic feature extraction}
Mean fundamental frequency is extracted using Praat via Parselmouth. Speaking 
rate is estimated from Whisper-medium word-level timestamps as
\[
r_{\text{spkrate}}
=
\frac{N_{\text{syllables}}}{T},
\]
where \(N_{\text{syllables}}\) denotes the total number of syllables and \(T\) 
represents the duration between the first and last detected word timestamps. 
Syllable counts are obtained using the CMU Pronouncing Dictionary (CMUdict) 
through the \texttt{pronouncing} package.

The voiced ratio is computed as
\[
r_{\text{voiced}}
=
\frac{N_{\text{voiced}}}{N_{\text{total}}},
\]
where \(N_{\text{voiced}}\) and \(N_{\text{total}}\) denote the number of voiced 
and total pitch frames, respectively.

The final prosodic feature vector is defined as
\[
\mathbf{p}
=
[
F0_{\text{mean}},
r_{\text{spkrate}},
r_{\text{voiced}}
].
\]

\paragraph{Audio-prosody fusion}
The prosodic feature vector \(\mathbf{p}\) is projected into a learned embedding 
space and concatenated with the utterance-level audio embedding. The combined 
representation is then passed through a linear projection layer to obtain the 
final speaker representation.

\subsection{Frame-level Aggregation}

While utterance embeddings provide a compact speaker representation, important speaker-discriminative cues may be lost during pooling. We thus investigate aggregation at the frame level. Specifically, frame representations are extracted from WavLM-ECAPA-TDNN before attentive statistical pooling (ASP) and aggregation is performed across multiple utterances prior to obtaining the final speaker embedding.

\subsubsection{Frame Concatenation} \label{audio_only_framepool}

Given frame sequences from $N$ utterances, we concatenate the frame representations along the temporal dimension prior to ASP, resulting in a unified sequence containing $\sum_{i=1}^{N} T_i$ frames. Speaker pooling is then applied over the aggregated sequence.

\subsubsection{Frame-level Audio-Text Fusion}\label{frame_level_multimodal}

Audio representations are obtained using the frame-level aggregation strategy described above. In parallel, LUAR is used to aggregate textual information across the same set of utterances, producing a speaker-level linguistic representation. The resulting audio and text embeddings are fused through concatenation followed by a linear projection.

We additionally investigate weighted fusion by assigning different weights to the audio and text embeddings prior to fusion. Specifically, we evaluate equal weighting using $(0.5,0.5)$ for audio and text embeddings, respectively, as well as a text-dominant setting using $(0.2,0.8)$. These configurations allow us to examine the relative contribution of linguistic and acoustic information to speaker verification performance. Since audio features typically encode strong speaker-discriminative features, forcing a text-dominant setting allows us to better assess the contribution of additional speaker information encoded in the linguistic content.
\subsubsection{Frame-level Audio Aggregation with Prosodic Fusion}
\label{frame_level_prosody}
While the acoustic branch aggregates frame-level representations across multiple
utterances, prosodic information is represented using utterance-level features.
For each utterance, we extract three prosodic features: mean fundamental
frequency ($F0_{\text{mean}}$), speaking rate ($r_{\text{spkrate}}$), and
voiced ratio ($r_{\text{voiced}}$), as described in
Section~\ref{multimodal_prosody}.
Given $N$ utterances from a speaker, the prosodic feature vectors are
aggregated by computing their mean and standard deviation across utterances,
yielding a six-dimensional statistical summary. This summary captures both the
overall prosodic characteristics of a speaker and their variation across
multiple utterances. The resulting prosodic representation is projected into a learned embedding space and fused with the speaker-level acoustic embedding
obtained from frame concatenation followed by attentive statistical pooling.
Fusion is performed through concatenation followed by a linear projection to
produce the final speaker representation.

\subsubsection{Hybrid Acoustic-Textual Speaker Verification}
\label{hybrid}

Existing multimodal architectures typically perform fusion using utterance-level 
representations derived independently from each modality. However, 
speaker-discriminative cues can be revealed not only by linguistic content but 
also by their acoustic realization. Previous studies have shown that speakers 
exhibit characteristic pronunciation and articulation patterns, even when 
producing the same lexical units \cite{nolan1945phonetic,laver1980phonetic}. 
We therefore hypothesize that modeling interactions between lexical tokens and 
frame-level acoustic representations enables the model to capture 
speaker-discriminative information that may otherwise be lost during 
utterance-level pooling.

\paragraph{Acoustic and textual representations}
Given an utterance, we first extract frame-level acoustic representations from 
WavLM-ECAPA-TDNN before attentive statistical pooling. Let 
\(H \in \mathbb{R}^{T \times C}\) denote the resulting acoustic feature map, 
where \(T\) is the number of acoustic frames and \(C\) is the acoustic feature 
dimension.

The corresponding ASR transcript is tokenized using the LUAR tokenizer to 
obtain a sequence of embeddings \(E \in \mathbb{R}^{L \times D}\), where 
\(L\) denotes the token sequence length after tokenization and padding, and 
\(D\) is the LUAR embedding dimension.\footnote{Padding and special tokens are masked.}

\paragraph{Token-to-frame cross-attention}
To align lexical and acoustic information, we use text tokens as queries and 
acoustic frames as keys and values
\[
Q = EW_Q,\qquad K = HW_K,\qquad V = HW_V,
\]
where \(Q \in \mathbb{R}^{L \times d_k}\), 
\(K \in \mathbb{R}^{T \times d_k}\), and 
\(V \in \mathbb{R}^{T \times d_v}\).

The token-to-frame attention matrix is computed as
\[
A = \mathrm{softmax}
\left(\frac{QK^\top}{\sqrt{d_k}}\right),
\]
where the softmax operation is applied over the acoustic frame dimension.
The resulting alignment matrix \(A \in \mathbb{R}^{L \times T}\) is used to 
compute token-conditioned acoustic representations:
\[
\tilde{V}=AV, \qquad \tilde{V}\in\mathbb{R}^{L\times d_v}.
\]

\paragraph{Hybrid representation learning}
To preserve both lexical and acoustic information, we concatenate the original 
token embeddings with the aligned acoustic representations:
\[
Z=[E||\tilde{V}].
\]
The concatenated representation is projected into a shared hybrid space
\[
Z'=ZW_f,
\]
where \(W_f\) is a learnable projection matrix and 
\(Z'\in\mathbb{R}^{L\times d_h}\).

The fused token sequence is then contextualized using a single-layer Transformer 
encoder:
\[
\hat{Z}=\mathrm{TransformerEncoder}(Z').
\]

An utterance-level hybrid representation is obtained through masked mean pooling
\[
z_{\mathrm{hyb}}
=
\frac{1}{|\mathcal{M}|}
\sum_{m\in\mathcal{M}}\hat{Z}_m,
\]
where \(\mathcal{M}\) denotes the set of non-padding and non-special token 
indices, resulting in \(z_{\mathrm{hyb}}\in\mathbb{R}^{d_h}\).

\paragraph{Speaker embedding generation}
In parallel, the full transcript is encoded using LUAR to obtain an 
utterance-level textual representation
\[
z_{\mathrm{luar}}
=
W_{\mathrm{luar}}\mathrm{LUAR}(x_{\mathrm{text}}),
\]
where \(W_{\mathrm{luar}}\) is a learnable projection matrix and 
\(z_{\mathrm{luar}}\in\mathbb{R}^{d_h}\).

Finally, the pooled hybrid representation and utterance-level LUAR 
representation are concatenated and projected to produce the final speaker 
embedding
\[
z_{\mathrm{spk}}
=
W_o[z_{\mathrm{hyb}}||z_{\mathrm{luar}}],
\]
where \(W_o\) is a learnable projection matrix.
\subsubsection{Recursive Joint Cross-Attention (RJCA)}
\label{rjca}

As an alternative multimodal fusion strategy, we investigate the Recursive Joint 
Cross-Attention (RJCA) framework proposed in \cite{praveen2024audio}. Unlike 
simple feature concatenation, RJCA explicitly models intra-modal and cross-modal 
relationships through recursive attention operations.

\paragraph{Utterance-level multimodal representations}
Given \(N\) utterances from a speaker, we extract a sequence of audio embeddings
\[
A=\{a_1,\ldots,a_N\}
\]
using WavLM-ECAPA-TDNN and a corresponding sequence of textual embeddings
\[
T=\{t_1,\ldots,t_N\}
\]
using LUAR.

\paragraph{Recursive joint cross-attention}
The audio and text embedding sequences are first processed using 
modality-specific BiLSTMs to capture dependencies across utterances within each 
modality. Following \cite{praveen2024audio}, the resulting audio and text 
representations are concatenated to construct a joint multimodal representation.

This joint representation is then used to compute cross-attention with each 
individual modality, allowing the model to capture both modality-specific 
dependencies and interactions between acoustic and textual speaker cues. We 
apply two recursive joint cross-attention layers, where the attended 
representations produced by one layer are provided as input to the next layer.

\paragraph{Aggregated textual representation}
Motivated by the observation that text-only speaker verification performance 
improves with increasing numbers of enrollment utterances~\cite{aggazzottigarg2026}, 
we further investigate an aggregated-text variant of RJCA.

Specifically, we first compute a speaker-level LUAR representation \(g\) using 
all available utterances and repeat this representation across utterance 
positions
\[
[(a_1,g),(a_2,g),\ldots,(a_N,g)].
\]

Since the textual representation already captures speaker-level information aggregated across all utterances, we omit the modality-specific BiLSTM for the text branch in this configuration.

\subsubsection{WavLM-Whisper cross-attention} 
To leverage complementary information captured by WavLM-ECAPA-TDNN and Whisper, we investigate a cross-attention-based fusion strategy operating on frame-level representations.

   Each utterance is first encoded independently and learnable positional embeddings are added to the frame-level features. Note that utterances per speaker are randomly sampled rather than from a continuous speech recording. So, only utterance-level positional embeddings are added. The frame-level features per utterance are then concatenated along the temporal dimension. To further contextualize the set of N utterances, we apply self attention across the aggregated sequence. Finally, cross-attention is applied with frame-level features from WavLM-ECAPA as query and frame-level features from Whisper as keys and values. This allows to further refine the speaker representation with complementary information from whisper. The resulting frame-level features are then aggregated using attentive statistical pooling. 
\section{Experiments}

\subsection{Data} 
\label{sec:data}
We use the Fisher English Training Speech Corpus \cite{cieri2004fisher}, a dataset of conversational telephone calls between two strangers with calls lasting up to 10 minutes. Existing datasets, such as LJSpeech \cite{ljspeech17} and LibriTTS \cite{zen2019libritts}, are read speech and thus not an indicator of an individual's linguistic choices. VoxCeleb \cite{nagrani2017voxceleb} and VoxCeleb2 \cite{chung2018voxceleb2} are also popular datasets for ASV, but only contain short clips of speech and thus often lack enough utterances per speaker for the text embeddings. Therefore, following previous work that successfully used LUAR representations on speech transcripts \cite{aggazzotti2024,aggazzotti2025asr}, we use the Fisher dataset for our experiments.\footnote{Fisher has the added benefit of being able to control for the topic under discussion. A topic-controlled evaluation produced similar results, though, so we choose the simpler setting with no topic control.}

We split the dataset by speaker, with 5,712 speakers in the training split, 250 speakers in the validation split, and 1,753 in the evaluation set.\footnote{For evaluation, we consider speakers who appear in at least two telephone calls, so that we can form disjoint enrollment and target utterance pools.} For model selection, we construct a fixed validation trial list with 100 target and 100 non-target trials per speaker. For each such speaker, we first construct separate enrollment and target datasets at the utterance level and then build multi-utterance trials by sampling from these pools.

We evaluate the proposed models under two training settings: end-to-end and two-stage training. In the end-to-end setting, all trainable components are optimized jointly. In the two-stage setting, the audio-only speaker verification baseline is first trained independently. The resulting audio encoder is then kept frozen, and the aggregation or fusion module is trained using the fixed audio representations. 
For two-stage training, the training speakers are partitioned into base-training (65\%) and fusion-training (35\%) subsets, while the validation speakers are split equally between the two stages. To construct multi-utterance evaluation trials, we use $N \in \{5, 10, 15\}$ utterances for both the enrollment and target sets. For both enrollment and target utterance pools, the subsets satisfy
$U_5 \subset U_{10} \subset U_{15}$, ensuring that any performance differences can be attributed to the inclusion of additional utterances rather than sampling variation.

We consider original-anonymized (O--A) and anonymized-anonymized (A--A) evaluation conditions following the VoicePrivacy evaluation framework \cite{tomashenko2024voiceprivacy}. In the O--A setting, enrollment utterances are taken from the original speech while target utterances are anonymized speech from a different call. In the A--A setting, both enrollment and target utterances are anonymized using the same method. For multimodal systems, Whisper-medium transcripts are generated from the corresponding audio used in each condition; thus, transcripts for anonymized utterances are obtained by transcribing the anonymized speech. Unlike in the O--A setting, where the original utterances of a target speaker are readily available for enrollment and compared to anonymized speech, the A--A setting assumes that the attacker has access to the anonymization system and compares the anonymized target samples to the anonymized enrollment utterances while avoiding the mismatch between the enrolled and target set.

To evaluate speaker verification systems under anonymization, all utterances requiring anonymization are processed using Stream-Voice-Anon \cite{kuzmin2026}. For each utterance to be anonymized, a target speaker is randomly selected from the LibriSpeech \cite{7178964} train-clean-360 and train-other-500 subsets, using only target utterances longer than 4 seconds. Thus, in the A--A condition, both enrollment and target utterances are anonymized independently using this procedure. Following \cite{kuzmin2026}, anonymization is performed using a fixed 2-frame delay and $\alpha=1$ for speaker embedding mixing.

\subsection{Baselines}\label{sec:baseline}
We consider two audio-only speaker verification baselines: an x-vector system \cite{snyder2018x} using a publicly available SpeechBrain pre-trained model trained on VoxCeleb1 and VoxCeleb2\footnote{\url{https://huggingface.co/speechbrain/spkrec-xvect-voxceleb}} and a WavLM-ECAPA-TDNN system. The WavLM-ECAPA-TDNN system utilizes ECAPA-TDNN \cite{desplanques2020ecapa} with WavLM \cite{chen2022wavlm} as the front-end feature extractor and is trained on the Fisher training split described in \autoref{sec:end_to_end}. ECAPA-TDNN extends the x-vector \cite{snyder2018x} framework and incorporates a Time Delay Neural Network (TDNN) layer together with a squeeze excitation (SE) block to model global channel dependencies. To perform multi-utterance evaluation, the baseline aggregates enrollment and target utterance embeddings using mean pooling. 

For the textual representation, we utilize LUAR~\cite{uar-emnlp2021}, a contrastively-trained authorship representation that captures speaker-identifying information from the linguistic style and content of transcribed speech \cite{aggazzotti2024,aggazzotti2025asr}. Specifically, we fine-tune LUAR on transcripts of the Fisher audio, which we transcribe using Whisper-medium,\footnote{\scriptsize{\url{https://github.com/openai/whisper}}} with speakers disjoint from the evaluation sets.\footnote{The Fisher Corpus includes human-transcribed transcripts of the audio, but to better represent a real-world scenario in which high-quality manual transcription is unavailable, we use automatic transcriptions (WER = $\sim$21\%).} The resulting transcriptions are then utilized to obtain textual representation: LUAR takes a list of utterances as input and outputs a single vector representation. 

\subsection{Setup}
\paragraph{End-to-end}\label{sec:end_to_end}
End-to-end models use a pretrained WavLM-Large frontend and a LUAR encoder fine-tuned on the full Fisher training set. The WavLM frontend remains fixed, while the ECAPA-TDNN backbone and aggregation or fusion modules are optimized jointly. For no aggregation baseline models, training uses a batch size of 64 utterances. For multi-utterance aggregation models, training uses batches of 64 speakers, where $N=5$ utterances are sampled for each speaker. All utterances are randomly cropped to 4 seconds. Models are optimized using AAM-Softmax (margin 0.2, scale 30) and AdamW with a learning rate of $5\times10^{-4}$, weight decay of $10^{-4}$, gradient clipping of 1.0, and a step scheduler with step size 5 and decay factor 0.5. The final speaker embedding dimension used is 192 for all models. Unless otherwise stated, end-to-end models are trained for 30 epochs.

The hybrid acoustic-textual ASV method (\autoref{hybrid}) is trained for 30 epochs using a learning rate of \(5\times10^{-5}\), weight decay of 0.01, gradient clipping of 1.0, and 4-head cross-attention with \(d_k=d_v=256\).
The LUAR encoder is fine-tuned on the Fisher training set with a learning rate of $2\times10^{-5}$, batch size 32, temperature 0.01, maximum token length 30, and 512-dimensional embeddings. Training is performed for 40 epochs with validation every 3 epochs. For the two-stage setup, LUAR is instead trained on the base-training speakers and used to initialize the corresponding multimodal models.

\paragraph{Two-stage training (lazy-informed)}
Two-stage models are initialized from the best audio-only model trained on the base-training speakers, as described in \autoref{sec:data}. The pretrained audio encoder and LUAR encoder trained on the base-training speakers remain frozen, and only the aggregation or fusion modules are optimized on the fusion-training speakers. The two-stage training setting is used for all utterance-level aggregation and multimodal models. Unless otherwise stated, models are trained for 5 epochs using a learning rate of \(10^{-4}\).

The query-attention aggregation model (\autoref{query_attention}) uses four attention heads and a query temperature of 0.3. RJCA (\autoref{rjca}) uses two recurrent layers, two cross-modal fusion layers, dropout 0.6, and four attention heads. Gradient clipping with a maximum norm of 1.0 is applied.

\paragraph{Two-stage training (semi-informed)}
To account for a stronger attacker with access to anonymized training data,
  we consider a semi-informed setting. Specifically, the Fisher training split is anonymized at the utterance level using the same Stream-Voice-Anon configuration employed for evaluation. Each semi-informed model variant is
  initialized from its corresponding lazy-informed model so that training
  starts from a representation learned on original utterances; in this setting,
  both the ECAPA backend and the aggregation module are further optimized. All
  semi-informed variants are trained with a learning rate of $10^{-4}$
  for 15 epochs.

\paragraph{Evaluation}
For each trial, enrollment and target utterances are aggregated according to the corresponding model architecture to obtain a single speaker embedding for each side. Speaker verification scores are computed using cosine similarity between the aggregated enrollment and target embeddings. Performance is reported as Equal Error Rate (EER).

\begin{figure}[t]
\includegraphics[width=1.02\columnwidth]{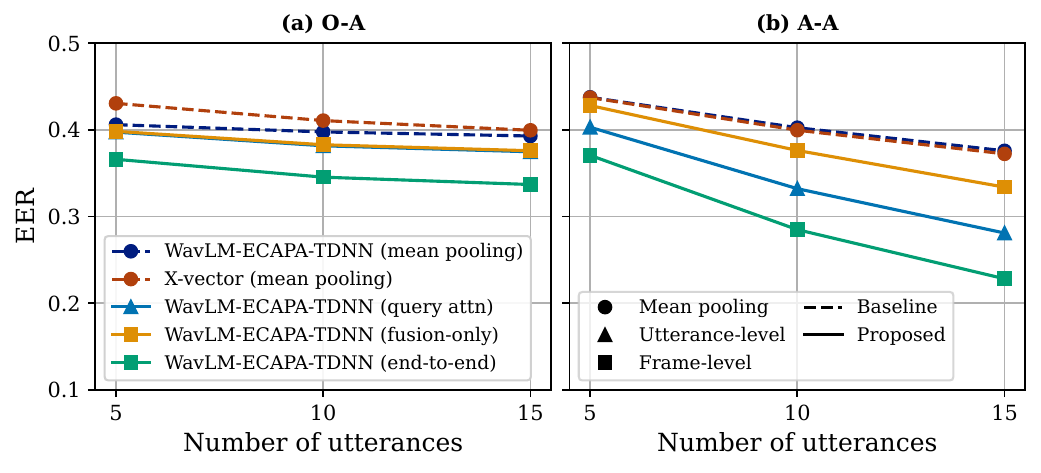}\hfill 
\caption{ASV performance (EER) across audio-only models on anonymized speech with aggregating utterances in the O--A (left) and A--A (right) settings. All models improve with more utterances, especially for A--A. Frame-level aggregation when trained end-to-end performs best in both settings.}
\label{fig:audio_aggr}
\end{figure}

\section{Results and Discussion}
\textbf{RQ1: Multi-utterance audio aggregation improves ASV performance on anonymized speech.}
\autoref{fig:audio_aggr} shows that both utterance-level and frame-level aggregation consistently outperform the WavLM-ECAPA-TDNN and x-vector baselines, across both O--A and A--A evaluation settings. In particular, frame-level aggregation provides larger gains, suggesting that speaker discriminative information preserved after anonymization is distributed across multiple temporal regions and can be exploited more effectively before utterance-level pooling. In addition, improvement gains in the A--A setting are more pronounced, where end-to-end frame aggregation results in a 39.30\% relative improvement at $N=15$ compared to the WavLM-ECAPA-TDNN baseline and a 25.24\% relative improvement over utterance-level aggregation. 

As shown in \autoref{semi-informed}, both aggregation strategies further benefit from semi-informed training, with frame-level aggregation consistently achieving the best performance across all evaluated enrollment conditions. When both enrollment and target speech are anonymized (A--A), the residual speaker characteristics introduced or preserved by the anonymization process are more consistent across both sides. Furthermore, the gains increase steadily with the number of available utterances, demonstrating that residual speaker information can accumulate across multiple anonymized recordings. 

Similar trends are observed in the O--O condition (\autoref{standard_asv}), indicating that the benefits of aggregation are not specific to anonymized speech. The persistence of aggregation gains under anonymization nevertheless demonstrates that anonymization does not completely remove speaker-discriminative information and that multi-utterance aggregation can effectively exploit residual identity cues even when only anonymized speech is available.

\begin{table}[t] 
\centering
\footnotesize
\caption{\textbf{Audio aggregation in the semi-informed setting.} EER (\% $\downarrow$) under A--A evaluation conditions. Semi-informed training reduces EER across all methods while preserving lazy-informed trends. Best results in each column are shown in bold.}
\begin{tabular}{llccc}
\hline
\textbf{Aggregation} & \textbf{Setting} & \textbf{N=5} & \textbf{N=10} & \textbf{N=15} \\
\hline
\multicolumn{5}{l}{\textbf{WavLM-ECAPA-TDNN Baseline}} \\
Mean Pooling & Lazy & 43.77 & 40.26 & 37.59 \\
             & Semi-Informed & 18.68 & 14.70 & 13.70 \\
\hline
\multicolumn{5}{l}{\textbf{Utterance-Level Aggregation}} \\
Query Attention & Lazy & 40.31 & 33.22 & 28.10 \\
                & Semi-Informed & 16.69 & 10.96 & 9.09\\
\hline
\multicolumn{5}{l}{\textbf{Frame-Level Aggregation}} \\
Frame Concat. (End-to-End) & Lazy & 37.07 & 28.49 & 22.84 \\
                          & Semi-Informed & \textbf{15.10} & \textbf{8.83} & \textbf{6.96} \\
\hline
\end{tabular}\label{semi-informed}
\end{table}

\begin{table}[t] 
\centering
\footnotesize
\caption{\textbf{Audio aggregation under O--O evaluation conditions.} EER (\% $\downarrow$) using a lazy-informed attacker (Trained on original speech only). Best results in each column are shown in bold.}
\begin{tabular}{llccc}
\hline
\textbf{Aggregation} & \textbf{Setting} & \textbf{N=5} & \textbf{N=10} & \textbf{N=15} \\
\hline
\multicolumn{5}{l}{\textbf{WavLM-ECAPA-TDNN Baseline}} \\
Mean Pooling & Lazy & 3.87 & 3.29 & 3.09 \\
\hline
\multicolumn{5}{l}{\textbf{Utterance-level Aggregation}} \\
Query Attention & Lazy & 3.39 & 2.54 & 2.27 \\
\multicolumn{5}{l}{\textbf{Frame-Level Aggregation}} \\
Frame Concat. (End-to-End) & Lazy & \textbf{3.26} & \textbf{2.33}& \textbf{2.10} \\
\hline
\end{tabular}\label{standard_asv}
\end{table}

\begin{table*}[t]
\centering
\begin{minipage}[t]{0.45\textwidth}
\footnotesize
\caption{\textbf{Unimodal and multimodal ASV performance.} EER (\% $\downarrow$) under O--A and A--A conditions, lazy-informed attacker (trained on original speech only). Best results in each column are shown in bold.}
\label{tab:multimodal_results}
\setlength{\tabcolsep}{3pt}
\begin{tabular}{llcccccc}
\hline
\multirow{2}{*}{\textbf{System}} &
\multirow{2}{*}{\textbf{Fusion/Aggr.}} &
\multicolumn{2}{c}{\textbf{N=5}} &
\multicolumn{2}{c}{\textbf{N=10}} &
\multicolumn{2}{c}{\textbf{N=15}} \\
\cline{3-8}
&
& \textbf{O--A} & \textbf{A--A}
& \textbf{O--A} & \textbf{A--A}
& \textbf{O--A} & \textbf{A--A} \\
\hline

\multicolumn{8}{l}{\textbf{Unimodal Systems}} \\
Audio-only & Mean Pooling
& 40.61 & 43.77
& 39.77 & 40.26
& 39.30 & 37.59 \\

Text-only & Learned aggr.
& 38.25 & 39.55
& 30.69 & 32.51
& 26.51 & 28.20 \\
\hline
\multicolumn{8}{l}{\textbf{Multi-modal Multi-Utterance Fusion}} \\
Audio + Prosody & Concat & 38.32 & 38.98 &36.30 &30.70 &35.54 &25.20  \\ 
Audio + Text & Concat
& 37.72 & 38.28
& 35.66 & 29.57
& 34.84 & 23.91 \\
Audio + Text & $0.5A + 0.5T$
& 36.79 & \textbf{37.18}
& 34.45 & \textbf{28.02}
& 33.51 & \textbf{22.63} \\

Audio + Text & $0.2A + 0.8T$
& \textbf{35.50} & 38.39
& \textbf{29.89} & 32.46
& \textbf{27.16} & 29.29  \\
\hline
\end{tabular}
\end{minipage}
\hfill 
\begin{minipage}[t]{0.47\linewidth}
\centering
\footnotesize
\caption{\textbf{A--A-only multimodal results.} EER (\% $\downarrow$) under A--A conditions using a lazy-informed attacker (trained only on original speech). Best results in each column are shown in bold.}
\label{tab:aa_only_results}
\setlength{\tabcolsep}{3pt}
\begin{tabular}{llccc}
\hline
\textbf{System} &
\textbf{Fusion / Aggreg.} &
\textbf{N=5} &
\textbf{N=10} &
\textbf{N=15} \\
\hline
\multicolumn{5}{l}{\textbf{Per-utterance Fusion}} \\
Hybrid Acoustic-Textual & Mean Pooling & 42.80 & 36.90 & 32.40 \\
Audio + Prosody & Mean Pooling & 41.88 & 35.74 & 31.09 \\
Audio + Text  & Mean Pooling & 41.25 & 34.70 & 29.82 \\
\hline
\multicolumn{5}{l}{\textbf{Multi-Utterance Fusion}} \\
WavLM--Whisper & Cross-Attn & 42.00 & 35.94 & 31.38 \\
\multirow{1}{*}{RJCA~\cite{praveen2024audio}}
& Audio + Global LUAR
& \textbf{40.15} & \textbf{32.35} & \textbf{27.23} \\
\hline
\end{tabular}
\end{minipage}
\end{table*} 

\textbf{RQ2: Multimodal systems outperform unimodal systems.}
\autoref{tab:multimodal_results} compares the audio-only (WavLM-ECAPA-TDNN baseline), text-only (LUAR), and multimodal systems under the anonymized speech evaluation setting. The text-only model consistently outperforms the audio-only baseline, demonstrating that speaker-specific lexical and linguistic patterns remain highly informative even after anonymization, particularly as more linguistic context becomes available across multiple utterances.

Combining audio with complementary modalities further improves performance. Both Audio+Prosody and Audio+Text outperform the audio-only baseline. Among the proposed models, the Audio+Text system with equal weighting ($0.5A+0.5T$) achieves the lowest A--A EER across all values of $N$, reducing the EER from 37.59\% to 22.63\% at $N=15$. Frame-level Audio+Prosody fusion also consistently improves over the audio-only baseline, reducing the A--A EER to 25.20\% at $N=15$. In contrast, the text-dominant weighting ($0.2A+0.8T$) achieves the best O--A performance across all values of $N$, reducing the EER to 27.16\% at $N=15$.

We further evaluate additional multimodal architectures under A--A conditions in \autoref{tab:aa_only_results}. When audio and complementary modalities are fused at the individual utterance level, both Audio+Prosody and Audio+Text continue to outperform the audio-only baseline, which is shown in \autoref{tab:multimodal_results}. At $N=15$, they relatively reduce the A--A EER by 20.7\% and 17.3\%, respectively, demonstrating that complementary modalities provide useful speaker information even with simple utterance-level fusion. Furthermore, the Hybrid Acoustic-Textual model (\autoref{hybrid}) also consistently outperforms the audio-only baseline despite performing token-level cross-modal fusion, relatively reducing the A--A EER by 2.22\%, 8.35\%, and 13.80\% for $N=5$, $10$, and $15$, respectively. RJCA further improves performance, achieving the lowest EERs across all values of $N$, indicating the benefit of modeling both intra-modal and inter-modal interactions between acoustic and linguistic representations. 

Overall, these results suggest that anonymization does not remove all speaker-discriminative information. Residual linguistic, prosodic, and cross-modal cues remain exploitable and combining complementary modalities consistently produces stronger privacy attacks than unimodal systems across a range of fusion strategies.

\textbf{RQ3: Frame-level aggregation is significantly better.}
Among the multimodal approaches, the proposed frame-level Audio+Text fusion achieves the lowest EERs across most evaluation settings (\autoref{tab:multimodal_results}). Overall, both the aggregation granularity and the incorporation of complementary modalities influence speaker verification performance, with frame-level multimodal aggregation providing the strongest privacy-leakage attack.

\section{Conclusions and Future Work}

In this work, we studied multi-utterance and multimodal speaker verification as an attack on speaker anonymization. We compared aggregation strategies operating at the utterance and frame levels and examined multimodal representations combining acoustic embeddings with linguistic content from ASR transcripts and prosodic features. Our results show that anonymization does not completely remove speaker-discriminative information: Aggregating audio across multiple anonymized utterances consistently improves verification performance, multimodal systems outperform their unimodal counterparts, and frame-level aggregation is the most effective strategy across settings. Linguistic cues prove particularly robust to voice anonymization, becoming increasingly informative as more utterances are available, and a semi-informed attacker with access to anonymized training data further amplifies these gains. These findings indicate that privacy evaluations based on isolated utterances and audio-only attack models may overestimate the protection offered by voice anonymization, and that comprehensive speaker privacy requires attention to linguistic and prosodic channels in addition to a speaker's vocal characteristics.

Our study also suggests directions for future work. Following the VoicePrivacy setup, anonymization was applied independently per utterance; a speaker-level scheme that assigns a consistent pseudo-speaker across utterances constitutes a complementary threat model whose interaction with multi-utterance aggregation merits investigation. In addition, our linguistic representations were derived from ASR transcripts of anonymized audio, and characterizing how transcription errors introduced by anonymization affect the linguistic modality would help isolate the contribution of each information source. Finally, extending the analysis to additional anonymization systems, languages, and speaking styles would further delineate the conditions under which residual speaker information can be exploited. We hope these findings inform the design of anonymization methods that jointly obscure acoustic, prosodic, and linguistic identity cues.

\section*{Acknowledgment}
This research is supported in part by the Office of the Director of National Intelligence (ODNI), Intelligence Advanced Research Projects Activity (IARPA), via the HIATUS Program contract \#D2022-2205150003 and the ARTS Program contract \#D2023-2308110001. The views and conclusions contained herein are those of the authors and should not be interpreted as necessarily representing the official policies, either expressed or implied, of ODNI, IARPA, or the U.S. Government. The U.S. Government is authorized to reproduce and distribute reprints for governmental purposes notwithstanding any copyright annotation therein.

\bibliographystyle{IEEEtran}
\bibliography{ref}

@inproceedings{praveen2024audio,
  title={Audio-visual person verification based on recursive fusion of joint cross-attention},
  author={Praveen, R Gnana and Alam, Jahangir},
  booktitle={2024 IEEE 18th International Conference on Automatic Face and Gesture Recognition (FG)},
  pages={1--5},
  year={2024},
  organization={IEEE},
  doi={10.1109/FG59268.2024.10582018}
}

@inproceedings{chen20253d,
  title={{3D-Speaker-Toolkit}: An open-source toolkit for multimodal speaker verification and diarization},
  author={Chen, Yafeng and Zheng, Siqi and Wang, Hui and Cheng, Luyao and Zhu, Tinglong and Huang, Rongjie and Deng, Chong and Chen, Qian and Zhang, Shiliang and Wang, Wen and others},
  booktitle={ICASSP 2025-2025 IEEE International Conference on Acoustics, Speech and Signal Processing (ICASSP)},
  pages={1--5},
  year={2025},
  organization={IEEE}
}

@inproceedings{krzywdziak2025merge,
  title={How to merge your embeddings: {S}tatistical vs attention-based speaker embedding aggregation for speaker verification with multiple enrollments},
  author={Krzywdziak, Justyna and Masztalski, Piotr and Romaniuk, Michal and Dudek, Milosz and Stepien, Joanna and Matuszewski, Mateusz and Hemmerling, Daria},
  booktitle={2025 33rd European Signal Processing Conference (EUSIPCO)},
  pages={26--30},
  year={2025},
  organization={IEEE}
}

@inproceedings{zeng2022attention,
  title={Attention back-end for automatic speaker verification with multiple enrollment utterances},
  author={Zeng, Chang and Wang, Xin and Cooper, Erica and Miao, Xiaoxiao and Yamagishi, Junichi},
  booktitle={ICASSP 2022-2022 IEEE International Conference on Acoustics, Speech and Signal Processing (ICASSP)},
  pages={6717--6721},
  year={2022},
  organization={IEEE}
}

@inproceedings{soleymani2018prosodic,
  title={Prosodic-enhanced siamese convolutional neural networks for cross-device text-independent speaker verification},
  author={Soleymani, Sobhan and Dabouei, Ali and Iranmanesh, Seyed Mehdi and Kazemi, Hadi and Dawson, Jeremy and Nasrabadi, Nasser M},
  booktitle={2018 IEEE 9th International Conference on Biometrics Theory, Applications and Systems (BTAS)},
  pages={1--7},
  year={2018},
  organization={IEEE}
}

@inproceedings{kuzmin2026,
author = {Kuzmin, Nikita and Liu, Songting and Lee, Kong Aik and Chng, Eng},
year = {2026},
booktitle={IEEE International Conference on Acoustics, Speech and Signal Processing (ICASSP)}, 
pages = {13587--13591},
title = {{Stream-Voice-Anon}: Enhancing utility of real-time speaker anonymization via neural audio codec and language models},
doi = {10.1109/ICASSP55912.2026.11463808}
}

@article{vandommelen1990,
author = {Van Dommelen, Wim},
year = {1990},
month = {07},
pages = {259-72},
title = {Acoustic Parameters in Human Speaker Recognition},
volume = {33 (Pt. 3)},
journal = {Language and Speech},
doi = {10.1177/002383099003300302}
}

@article{shriberg2005modeling,
  title={Modeling prosodic feature sequences for speaker recognition},
  author={Shriberg, Elizabeth and Ferrer, Luciana and Kajarekar, Sachin and Venkataraman, Anand and Stolcke, Andreas},
  journal={Speech Communication},
  volume={46},
  number={3-4},
  pages={455--472},
  year={2005},
  publisher={Elsevier}
}

@inproceedings{mary2006prosodic,
  title={Prosodic features for speaker verification.},
  author={Mary, Leena and Yegnanarayana, B},
  booktitle={Proceedings of Interspeech},
  number={17-21},
  pages={917--920},
  year={2006}
}

@inproceedings{shamsabadi2022differentially,
  title={Differentially private speaker anonymization},
  author={Shamsabadi, Ali Shahin and Srivastava, Brij Mohan Lal and Bellet, Aur{\'e}lien and Vauquier, Nathalie and Vincent, Emmanuel and Maouche, Mohamed and Tommasi, Marc and Papernot, Nicolas},
  booktitle={Proceedings on Privacy Enhancing Technologies Symposium},
  year={2023},
  number={1},
  pages={98--114},
  doi={https://doi.org/10.56553/popets-2023-0007}
}

@inproceedings{snyder2018x,
  title={X-vectors: Robust {DNN} embeddings for speaker recognition},
  author={Snyder, David and Garcia-Romero, Daniel and Sell, Gregory and Povey, Daniel and Khudanpur, Sanjeev},
  booktitle={IEEE International Conference on Acoustics, Speech and Signal Processing (ICASSP)},
  pages={5329--5333},
  year={2018},
  organization={IEEE}
}

@inproceedings{cieri2004fisher,
  title={{The {F}isher {C}orpus: A resource for the next generations of speech-to-text}},
  author={Cieri, Christopher and Miller, David and Walker, Karen},
  booktitle={Proceedings of the 4th International Conference on Language Resources and Evaluation},
  pages={69--71},
  year={2004}
}

@article{desplanques2020ecapa,
  title={{ECAPA-TDNN: Emphasized Channel Attention, Propagation and Aggregation in TDNN based speaker verification}},
  author={Desplanques, Brecht and Thienpondt, Jenthe and Demuynck, Kris},
  journal={Proceedings of Interspeech},
  doi = {10.21437/Interspeech},
  year={2020},
  pages={3830--3834}
}

@article{chen2022wavlm,
  title={{WavLM}: Large-scale self-supervised pre-training for full stack speech processing},
  author={Chen, Sanyuan and Wang, Chengyi and Chen, Zhengyang and Wu, Yu and Liu, Shujie and Chen, Zhuo and others},
  journal={IEEE Journal of Selected Topics in Signal Processing},
  volume={16},
  number={6},
  pages={1505--1518},
  year={2022},
  publisher={IEEE}
}

@inproceedings{uar-emnlp2021,
  author    = {Rafael A. Rivera Soto and Olivia Miano and Juanita Ordonez and Barry Chen and Aleem Khan and Marcus Bishop and Nicholas Andrews},
  title     = {Learning Universal Authorship Representations},
  booktitle = {Proceedings of the 2021 Conference on Empirical Methods in Natural Language Processing},
  year      = {2021},
  publisher = "Association for Computational Linguistics",
    url = "https://aclanthology.org/2021.emnlp-main.70",
    pages = "913--919",
}

@INPROCEEDINGS{aggazzottigarg2026,
  author={Aggazzotti, Cristina and Garg, Ashi and Cai, Zexin and Andrews, Nicholas},
  booktitle={IEEE International Conference on Acoustics, Speech and Signal Processing (ICASSP)}, 
  title={Content Anonymization for Privacy in Long-Form Audio}, 
  year={2026},
  volume={},
  number={},
  pages={13412-13416},
  doi={10.1109/ICASSP55912.2026.11461508}
}

@article{tomashenko2024voiceprivacy,
  title={{The VoicePrivacy 2024 challenge evaluation plan}},
  author={Tomashenko, Natalia and Miao, Xiaoxiao and Champion, Pierre and Meyer, Sarina and Wang, Xin and Vincent, Emmanuel and Panariello, Michele and Evans, Nicholas and Yamagishi, Junichi and Todisco, Massimiliano},
  journal={arXiv:2404.02677},
  year={2024}
}

@article{aggazzotti2024,
    author = {Aggazzotti, Cristina and Andrews, Nicholas and Smith, Elizabeth Allyn},
    journal = {Trans. of the Association for Computational Linguistics},
    pages = {875--891},
    title = {{Can Authorship Attribution Models Distinguish Speakers in Speech Transcripts?}},
    volume = {12},
    year = {2024}
}

@article{aggazzotti2025asr,
    author = {Aggazzotti, Cristina and Wiesner, Matthew and Smith, Elizabeth Allyn and Andrews, Nicholas},
    title = {{The Impact of Automatic Speech Transcription on Speaker Attribution}},
    journal = {Trans. of the Association for Computational Linguistics},
    volume = {13},
    pages = {1578-1596},
    year = {2025},
}

@misc{mehlman2025rhythm,
      title={Rhythm Features for Speaker Identification}, 
      author={Nick Mehlman and Thomas Thebaud and Dani Byrd and Shri Narayanan},
      year={2025},
      eprint={2506.06834},
      archivePrefix={arXiv},
      primaryClass={eess.AS},
      url={https://arxiv.org/abs/2506.06834}, 
}

@misc{ljspeech17,
  author       = {Keith Ito and Linda Johnson},
  title        = {The {LJ} Speech Dataset},
  howpublished = {\url{https://keithito.com/LJ-Speech-Dataset/}},
  year         = 2017
}

@inproceedings{zen2019libritts,
  title = {{LibriTTS}: A Corpus Derived from {LibriSpeech} for Text-to-Speech},
  author = {H. Zen and V. Dang and R. Clark and Y. Zhang and R. J. Weiss and Y. Jia and Z. Chen and Y. Wu},
  booktitle = {Proceedings of Interspeech},
  year = {2019},
  pages = {1526--1530},
  doi = {10.21437/Interspeech.2019-2441}
}

@inproceedings{nagrani2017voxceleb,
	author       = "Nagrani, A. and Chung, J.~S. and Zisserman, A.",
	title        = "{VoxCele}b: A large-scale speaker identification dataset",
	booktitle    = "Proceedings of Interspeech",
	year         = "2017",
}

@inproceedings{chung2018voxceleb2,
  title     = {{{VoxCeleb2}: Deep Speaker Recognition}},
  author    = {Joon Son Chung and Arsha Nagrani and Andrew Zisserman},
  booktitle = {Proceedings of Interspeech},
  pages     = {1086--1090},
  year = {2018}
}

@INPROCEEDINGS{7178964,
  author={Panayotov, Vassil and Chen, Guoguo and Povey, Daniel and Khudanpur, Sanjeev},
  booktitle={2015 IEEE International Conference on Acoustics, Speech and Signal Processing (ICASSP)}, 
  title={{Librispeech}: An {ASR} corpus based on public domain audio books}, 
  year={2015},
  volume={},
  number={},
  pages={5206-5210},
  doi={10.1109/ICASSP.2015.7178964}}

@article{nolan1945phonetic,
  title={The Phonetic Bases of Speaker Recognition},
  author={Nolan, Francis},
  journal={American history},
  volume={1861},
  number={1900},
  year={1945}
}

@book{laver1980phonetic,
  title={The phonetic description of voice quality},
  author={Laver, John},
  year={1980},
  publisher={Cambdrige University Press}
}
\end{document}